\documentclass[aps,prx,twocolumn,amsmath,amssymb, superscriptaddress,notitlepage,longbibliography]{revtex4-1}
\usepackage[colorlinks=true,linkcolor=blue,anchorcolor=red,citecolor=blue, urlcolor=blue]{hyperref}
\usepackage{bm}
\usepackage{graphicx}
\usepackage{epstopdf}
\usepackage{color}
\usepackage{cases}
\usepackage{amssymb}
\usepackage{amsmath}
\usepackage{subfigure}
\usepackage{enumerate}
\usepackage{longtable}

\begin{document}

\title{Theory for the negative longitudinal magnetoresistance in the quantum limit of Kramers Weyl semimetals}

\author{Bo Wan}
\affiliation{School of Physics, Nanjing University, Nanjing 210093, China}
\affiliation{Shenzhen Institute for Quantum Science and Technology and Department of Physics, Southern University of Science and Technology, Shenzhen 518055, China}

\author{Frank Schindler}
\affiliation{Department of Physics, University of Zurich, Winterthurerstrasse 190, 8057 Zurich, Switzerland}

\author{Ke Wang}
\affiliation{Shenzhen Institute for Quantum Science and Technology and Department of Physics, Southern University of Science and Technology, Shenzhen 518055, China}

\author{Kai Wu}
\affiliation{XTAL Incorporated, San Jose, CA 95112, USA}

\author{Xiangang Wan}
\affiliation{School of Physics, Nanjing University, Nanjing 210093, China}

\author{Titus Neupert}
\affiliation{Department of Physics, University of Zurich, Winterthurerstrasse 190, 8057 Zurich, Switzerland}

\author{Hai-Zhou Lu}\email{luhz@sustech.edu.cn}
\affiliation{Shenzhen Institute for Quantum Science and Technology and Department of Physics, Southern University of Science and Technology, Shenzhen 518055, China}
\affiliation{Shenzhen Key Laboratory of Quantum Science and Engineering, Shenzhen 518055, China}


\date{\today}

\begin{abstract}
Negative magnetoresistance is rare in non-magnetic materials.
Recently, a negative magnetoresistance has been observed in the quantum limit of $\beta$-Ag$_2$Se, where only one band of Landau levels is occupied in a strong magnetic field parallel to the applied current. $\beta$-Ag$_2$Se is a material that host a Kramers Weyl cone with band degeneracy near the Fermi energy. Kramers Weyl cones exist at time-reversal invariant momenta in all symmorphic chiral crystals, and at a subset of these momenta, including the $\Gamma$ point, in non-symmorphic chiral crystals.
Here, we present a theory for the negative magnetoresistance in the quantum limit of Kramers Weyl semimetals. We show that, although there is a band touching similar to those in Weyl semimetals, negative magnetoresistance can exist without a chiral anomaly. We find that it requires screened Coulomb scattering potentials between electrons and impurities, which is naturally the case in $\beta$-Ag$_2$Se.
\end{abstract}


\maketitle


\section{Introduction}

Magnetoresistance measures the change of the electric resistance of a solid due to magnetic fields. In magnetic materials, randomly-oriented ferromagnetic domains suppress the tunneling of electrons and increase the resistance. A magnetic field can align the domains and thus lower the resistance, leading to a negative magnetoresistance. In contrast, negative magnetoresistance is rare in non-magnetic materials \cite{Argyres56prb}, because the Lorentz force imposed by the magnetic field prevents electrons from moving forward.
One of the mechanisms of negative magnetoresistance is weak localization \cite{Lee85rmp}, which is induced by quantum interference and thus only survives at extremely low temperatures. Recently, negative magnetoresistance at higher temperatures in non-magnetic topological insulators \cite{Wang12nr,He13apl,Wiedmann16prb,Wang15ns,Breunig17nc,Assaf17prl} and semimetals \cite{Kim13prl,Kim14prb,Li16np,ZhangCL16nc,HuangXC15prx,Xiong15sci,LiCZ15nc,ZhangC17nc,LiH16nc,Arnold16nc,YangXJ15arXiv,YangXJ15arXiv-NbAs,WangHC16prb} has attracted tremendous interest.
In topological semimetals, the negative magnetoresistance is widely believed to be interpretable as a manifestation of the chiral anomaly, that is, the violation of chiral symmetry by quantum effects \cite{Adler69pr,Bell69Jackiw,Nielsen81npb}.
In topological insulators, the negative magnetoresistance is found to be related to the anomalous velocity induced by a nontrivial distribution of Berry curvature \cite{Dai17prl}. Because of the nontrivial mechanism behind each of the cases, a negative magnetoresistance observed in novel systems warrants a detailed study.

In a recent experiment on a single-crystalline silver chalcogenide material $\beta$-Ag$_2$Se \cite{ZhangCL17prb_Ag2Se}, a negative magnetoresistance has been observed when the magnetic field is parallel to the current.
$\beta$-Ag$_2$Se is among the material candidates for a class of systems called Kramers
Weyl semimetals, in which Weyl nodes are
pinned to time-reversal invariant momenta (TRIMs) in
the Brillouin zone because of time-reversal symmetry \cite{Bradlyn16sci,ChangGQ18nmat}.
The Kramers Weyl nodes generically appear in all chiral crystals, i.e., is crystals that lack any roto-inversion symmetries and thus have a sense of handedness \cite{ChangGQ18nmat}.
Specifically, in symmorphic chiral crystals, every Kramers
pair of bands at every TRIM is guaranteed to host a Weyl
cone; while in non-symmorphic chiral crystals, it is
true for a subset of TRIMs only, which however always
includes the $\Gamma$ ($\mathbf{k}=0$) point. $\beta$-Ag$_2$Se provides an instance of the latter case.
In $\beta$-Ag$_2$Se, a negative magnetoresistance of about $-20\%$ was observed at a magnetic field of about 9~T. At such a strong magnetic field, the system has entered the quantum limit, i.e., only the lowest Landau band crosses the Fermi energy. In the quantum limit, the magnetoresistance depends subtly on scattering mechanisms \cite{Lu15prb-QL,Goswami15prb,ZhangSB16njp}. Besides, since the Kramers Weyl nodes are protected from symmetries in chiral space groups, it provides a new platform for investigating the scattering mechanisms.

In this work, we present a theory for the longitudinal magnetoresistance in the quantum limit of a Kramers Weyl semimetal in strong parallel magnetic fields.
We start with a generic model with one Kramers Weyl cone and use the standard Kubo formalism to calculate the conductivity, considering impurity scattering with screened Coulomb potentials and Gaussian potentials.
We show that in the quantum limit the resistance has a $1/B$ dependence in the presence of impurities with screened Coulomb potential \cite{Dubinskaya69jetp}, and thus indeed gives rise to a negative magnetoresistance. In many Weyl semimetals that emerge from band inversion, Weyl nodes of opposite chirality are degenerate in energy, e.g. due to some mirror symmetry. This is generically not the case for the Kramers Weyl nodes in chiral crystals, so the inter-valley charge pumping and relaxation is absent in our calculations, suggesting that the negative magnetoresistance in the Weyl semimetal can exist without any apparent link to the chiral anomaly.
Although magnetoresistance in the quantum limit has been systematically studied for a number of models and potentials \cite{Lu15prb-QL,Goswami15prb,ZhangSB16njp}, the case we study here has not been addressed before.

To justify our results, we will present the step-by-step details of the calculation.
The paper is organized as follows. In Sec.~\ref{Sec:Model}, we introduce a generic model for Kramers Weyl semimetals. In Secs.~\ref{Sec:Time} and~\ref{Sec:Conductivity}, we present the calculations for the transport time in the presence of the screened Coulomb scattering potential and longitudinal conductivity in the quantum limit, respectively. To justify the isotropic model and screened Coulomb scattering potential we used, we discuss the effect of anisotropy and Gaussian scattering potential in Secs. \ref{Sec:Gaussian} and \ref{Sec:Anisotropy}, respectively.
Finally, we summarize and discuss  the results in Sec.~\ref{Sec:Discussion}.

\section{Model and Landau Bands} \label{Sec:Model}
A Kramers Weyl cone can be described by the effective $\boldsymbol{k}\cdot\boldsymbol{p}$ model proposed in \cite{ZhangCL17prb_Ag2Se},
 \begin{eqnarray} \label{model}
 \mathcal{H}=u (k_x^2+k_y^2+k_z^2)+v (k_x\sigma_x+k_y\sigma_y+k_z\sigma_z),
  \end{eqnarray}
where we have suppressed the anisotropy of the model parameters $u$ and $v$ present in the original model. We will find that this isotropic model can capture the main physics of negative magnetoresistance. In contrast to the band-inversion Weyl cone, the parabolic term overwhelms the linear term and the energy difference between paired Weyl nodes can be much larger than the temperature scale in Kramers Weyl semimetals. It is reasonable to consider only one Weyl cone as the longitudinal conductivity is dominated by electrons near the Fermi surface.

\begin{figure}[tbph]
\centering \includegraphics[width=0.45\textwidth]{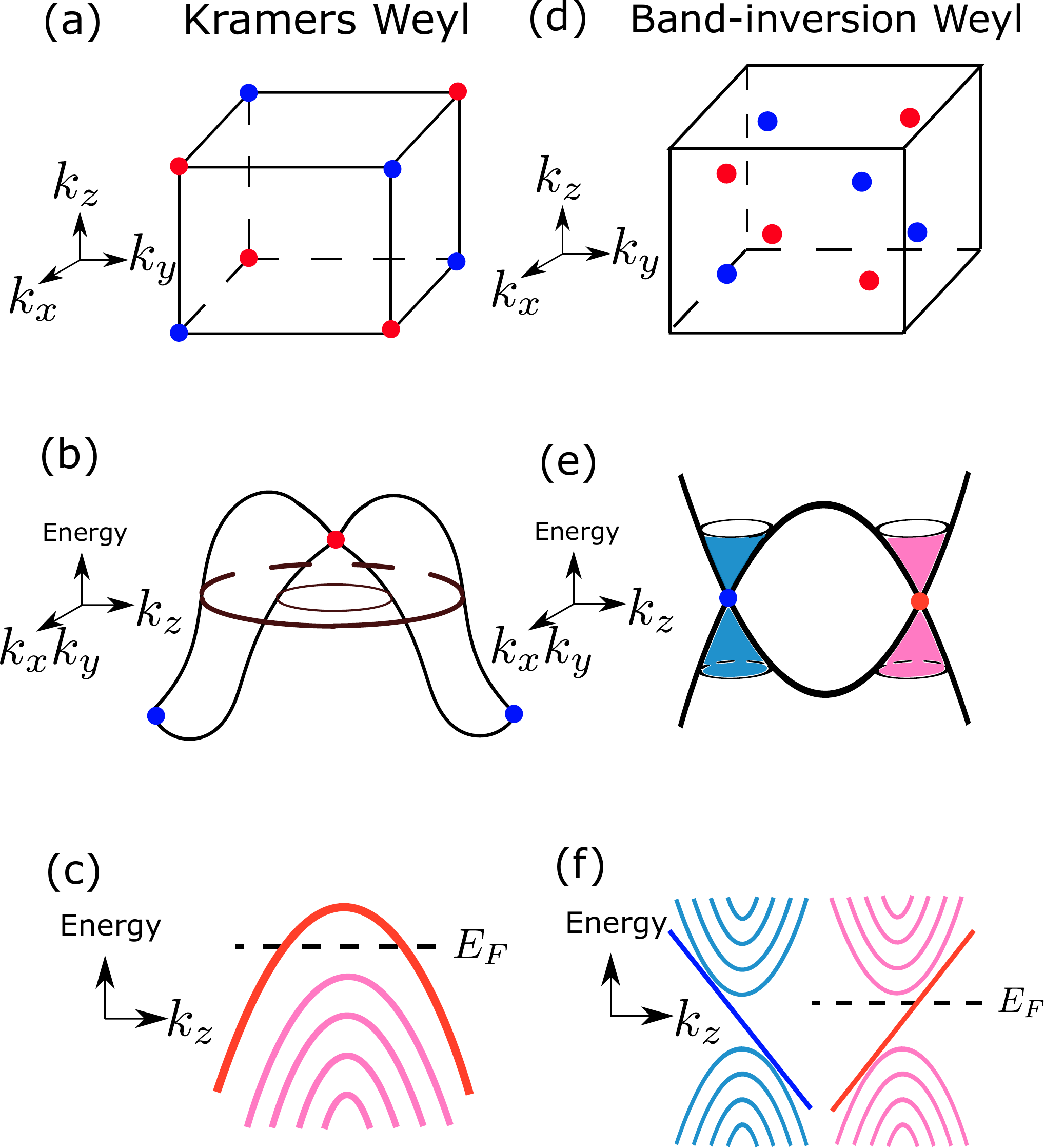}
\caption{Schematics of Kramers [(a)-(c)] and band-inversion induced [(d)-(f)] Weyl semimetals. (a) The Weyl nodes in a Kramers Weyl semimetal are located at the time-reversal invariant momenta in the Brillouin zone. (b) The energy dispersion in the vicinity of the Kramers Weyl nodes. (c) The Landau bands for the Weyl cone of a Kramers node [red in (b)] in a strong magnetic field. The dashed line indicates the position of the Fermi energy $E_{\mathrm{F}}$. (d) The Weyl nodes in a band-inversion Weyl semimetal are located somewhere between the time-reversal invariant momenta. (e) The energy dispersion near the band-inversion Weyl nodes. (f) The Landau levels for (e) in a strong magnetic field.}
\label{Fig:Band}
\end{figure}

The energy spectrum of this model has two bands
\begin{eqnarray}
 \mathcal{E}_{\pm}(\boldsymbol{k})=u(k_x^2+k_y^2+k_z^2)\pm v\sqrt{k_x^2+k_y^2+k_z^2},
\end{eqnarray}
which are schematically shown in Fig.~\ref{Fig:Band}~(b) in the plane $k_x=k_y=0$. At $\boldsymbol{k}=0$, a Kramers Weyl node forms as band $\mathcal{E}_+$ touches band $\mathcal{E}_-$. A similar Weyl cone is also present in BiTeI \cite{Murakami03sci}.
Weyl nodes always come in pairs because of the fermion-doubling theorem \cite{Nielsen83plb}.
For Weyl semimetals in which these pairs are (nearly) degenerate in energy, charges can be pumped from one Weyl cone to another Weyl cone of the opposite chirality in an external electric field, and both inter- and intra-cone scattering has to be considered. However, as this kind of energetic degeneracy of Weyl cones is not found in Kramers Weyl semimetals, we can assume that the intra-node scattering dominates. Later, we will show that a negative magnetoresistance can arise in this one-node Kramers Weyl semimetal in the presence of impurities with a screened Coulomb scattering potential.

In a magnetic field, the energy spectrum is quantized into a set of one-dimensional (1D) bands of Landau levels. Because the present model is isotropic, we can assume that the magnetic field is applied along the $z$ direction, that is, $\boldsymbol{B}=(0,0,B)$. Along the direction of the magnetic field the wave vector $k_z$ is a good quantum number, so the 1D Landau bands disperse with $k_z$.  We adopt the Landau gauge in which the vector potential is $\boldsymbol{A}=(-yB,0,0)$. Under the Peierls replacement, the wave vector becomes $\boldsymbol{k} \rightarrow (k_x- eyB/\hbar,-i\partial_y,k_z)$, and
the Hamiltonian in the applied magnetic field becomes
 \begin{eqnarray}\label{HamwithMag}
 \mathcal{H}=u\sum_{i=1}^{3}(k_i+eA_i)^2+v\sum_{i=1}^{3}(k_i+eA_i)\sigma_i. \end{eqnarray}
The chosen vector potential breaks the translational symmetry along the $y$-direction but not along the $x$- or $z$-direction, $k_x$ and $k_z$ are still good quantum numbers. Introducing the ladder operators \cite{Shen04prl,Shen04prbrc}
\begin{eqnarray}
a&\equiv & -[(y-\ell_B^2k_x)/\ell_B+\ell_B \partial_y]/\sqrt{2}, \\
a^{\dagger} &\equiv & -[(y-\ell_B^2k_x)/\ell_B-\ell_B \partial_y]/\sqrt{2},
\end{eqnarray}
we can replace the wave vectors $k_x^2+k_y^2 \rightarrow \omega(a^{\dagger}a+1/2)$, $k_+ \rightarrow (\sqrt{2}/\ell_B)a^{\dagger}$, $k_- \rightarrow (\sqrt{2}/\ell_B)a$, where $\ell_B=\sqrt{\hbar/eB}$ is the magnetic length, $\omega=2u/\ell_B^2$, and we have defined $k_\pm = k_x \pm i k_y$. Then the Hamiltonian in Eq.~(\ref{HamwithMag}) becomes
\begin{equation}
\mathcal{H}=
\begin{bmatrix}
\mathcal{M}^+_a&\eta a\\
\eta a^{\dagger}&\mathcal{M}^-_a
\end{bmatrix},
\end{equation}
where $\mathcal{M}^{\pm}_a=\omega(a^{\dagger}a+1/2)+uk_z^2\pm v k_z$ and $\eta=\sqrt{2}v/\ell_B$. With the trial wave functions $(c_1|\nu-1 \rangle,c_2 | \nu\rangle)^\mathsf{T}$ for $\nu \geq1$, and $(0,|0\rangle)^\mathsf{T}$ for $\nu=0$, where $\nu$ indexes the Hermite polynomial wave functions given below, the eigenenergies $E$ can be obtained from the secular equation
\begin{equation}
\begin{aligned}
\mbox{det}\begin{bmatrix}
\mathcal{M}^+_{\nu}-E&\eta \sqrt{\nu}\\
\eta \sqrt{\nu}&\mathcal{M}^-_{\nu}-E
\end{bmatrix}=0,
\end{aligned}
\end{equation}
for $\nu \geq 1$; and $\mathcal{M}^-_{\nu}-E=0$ for $\nu=0$, where $\mathcal{M}^{\pm}_{\nu}=\omega\nu\mp \omega/2+uk_z^2\pm vk_z$. The eigenenergies are found as
\begin{eqnarray}\label{Eq:Ek}
E_{\nu}^{\pm}(k_z)=
\begin{cases}
\mathcal{M} \pm \sqrt{(k_zv-\omega/2)^2+\eta^2 \nu}, & \mbox{  for $\nu \geq 1$},\\
u(k_z-v/2u)^2+u/\ell_B^2-v^2/4u, & \mbox{  for $\nu=0$},
\end{cases} \nonumber \\
\end{eqnarray}
where $\mathcal{M}=\omega \nu+k_z^2u$. This gives the spectrum of a set of Landau energy bands ($\nu$ as band index), which is shown in Fig.~\ref{Fig:Band}~(c). The eigenstates found for $\nu \geq 1$ are
\begin{equation}
\begin{aligned}
|\nu,k_x,k_z,+ \rangle &=& \begin{bmatrix}
\cos(\theta^{\nu}_{k_z}/2) |\nu-1\rangle \\
\sin(\theta^{\nu}_{k_z}/2) |\nu\rangle
\end{bmatrix} | k_x,k_z\rangle, \\
|\nu,k_x,k_z,- \rangle &=& \begin{bmatrix}
\sin(\theta^{\nu}_{k_z}/2) |\nu-1\rangle \\
-\cos(\theta^{\nu}_{k_z}/2) |\nu\rangle
\end{bmatrix} | k_x,k_z\rangle,\\
\end{aligned}
\end{equation}
and the eigenstate for $\nu=0$ is
\begin{equation}
|\nu=0,k_x,k_z\rangle =\begin{bmatrix}
0 \\ |0\rangle
\end{bmatrix}|k_x,k_z \rangle,
\end{equation}
where $\cos(\theta^\nu_{k_z}/2)= (k_zv-\omega/2)/\sqrt{(k_zv-\omega/2)^2+\eta^2\nu}$. The wave functions in real-space $\psi_{\nu,k_x,k_z}(\mathbf{r}) \equiv \langle \mathbf{r} | \nu,k_x,k_z\rangle$ are found to be
\begin{equation}
\begin{aligned}
\psi^{\nu}_{k_x,k_z}(\mathbf{r})
= \frac{C_{\nu}}{\sqrt{L_xL_z\ell_B}}e^{ik_xx}e^{ik_zz}e^{-\xi^2/2} \mathcal{H}_{\nu}(\xi),
\end{aligned}
\end{equation}
for $\nu \geq 1$
and
\begin{equation}
\psi^{0}_{k_x,k_z}(\mathbf{r})
= \frac{1}{\sqrt{L_xL_z\ell_B\sqrt{\pi}}}e^{ik_xx}e^{ik_zz}e^{-\xi^2/2},
\end{equation}
for $\nu=0$.
Here, $C_{\nu}=1/\sqrt{\nu!2^{\nu}\sqrt{\pi}}$ , $L_x$ and $L_z$ are the lengths of the sample in the $x$ and $z$ directions, respectively, $\xi=(y-y_0)/\ell_B$, $y_0=k_x \ell_B^2$ is the guiding center and $\mathcal{H}_{\nu}(\xi)$ are the Hermite polynomials. $k_x$ does not appear in the energy spectrum explicitly, because different $k_x$ give rise to the Landau degeneracy $N_{\mathrm{L}}=1/2\pi\ell_B^2$ per unit area in the x-y plane.

In what follows, we only concentrate on the quantum limit, which means only the lowest ($\nu=0$) Landau band crosses the Fermi energy [see Fig. \ref{Fig:Band} (b)].
The analytical solution of the Landau bands allows us to determine the analytical solution to the longitudinal magnetoresistance in the quantum limit of the Kramers Weyl cone.

\section{Screened Coulomb Scattering Potential and Transport Time} \label{Sec:Time}

To calculate the longitudinal conductivity, we need to know the transport time of carriers \cite{ZhangSB16njp,Goswami15prb}, which sensitively depends on the impurity scattering potential.
Following Abrikosov's treatment when exploring the linear magnetoresistance in the quantum limit of Ag$_2$Se under perpendicular magnetic fields \cite{Abrikosov98prb}, we adopt a screened Coulomb scattering potential for the impurities,
\begin{eqnarray}\label{Ur}
U(\mathbf{r})=\sum_{i}U(\mathbf{r}-\mathbf{R}_i),
\end{eqnarray}
with
\begin{eqnarray}\label{Ur2}
U(\mathbf{r}-\mathbf{R}_i)= \frac{e^2}{4\pi \epsilon |\mathbf{r}-\mathbf{R}_i|} e^{-\kappa |\mathbf{r}-\mathbf{R}_i|},
\end{eqnarray}
where $\epsilon$ is the dielectric constant, $e$ is the electron charge, and $1/\kappa$ is the screening length. The screened Coulomb potential is justified because excess silver atoms form clusters doping the rest of the material\cite{Abrikosov98prb}.
This potential is different from the ionic potential under the Thomas-Fermi approximation \cite{Goswami15prb}. Following the standard random phase approximation \cite{Abrikosov98prb,Bruus04book},
\begin{eqnarray}\label{length}
\kappa^2&=&-\frac{e^2}{\epsilon }\frac{1}{2\pi \ell^2_B} \frac{1}{\beta} \nonumber\\
&& \times  \sum_{m} \int_{-\infty}^{+\infty}\frac{dk_z}{2\pi}\frac{1}{(iw_m+E_{\mathrm{F}}-E^0_{k_z})^2},
\end{eqnarray}
where $\beta=1/{k_B T}$, $w_m=(2m+1)\pi/\beta$ are the Matsubara frequencies of fermions, $E_{\mathrm{F}}$ is the Fermi energy, and the energy spectrum $E^0_{k_z}$ of the $\nu=0$ Landau band has been given in Eq.~(\ref{Eq:Ek}). After summing over the Matsubara frequencies, we have
\begin{eqnarray}
\frac{1}{\beta} \sum_m \frac{1}{(iw_m+E_{\mathrm{F}}-E^0_{k_z})^2} = \frac{\partial n_{\mathrm{F}} (E_{\mathrm{F}}-E^0_{k_z})}{\partial (E_{\mathrm{F}}-E^0_{k_z})}.
\end{eqnarray}
Substituting this into Eq.~(\ref{length}), we have
\begin{eqnarray}\label{kappa}
\kappa^2 = \frac{e^2}{2\pi^2 \ell^2_B \epsilon \hbar v_{\mathrm{F}}^0},
\end{eqnarray}
where $v_{\mathrm{F}}^0$ is the Fermi velocity of the $\nu=0$ Landau band. To calculate the transport time, we need to calculate the scattering matrix elements,
\begin{eqnarray}
U_{\nu,\nu'} \equiv \langle \nu, k_x,k_z | U(\mathbf{r}) | \nu',k'_x,k'_z \rangle,
\end{eqnarray}
where $U(\mathbf{r})$ is defined in Eq.~(\ref{Ur}), and $\nu, \nu'$ are indices for the Landau bands. In the quantum limit, due to the strong magnetic field, the energy spectrum has split into 1D Landau bands, and the spacing between the Landau bands is large. Moreover, because the Fermi energy crosses only the $\nu=0$ Landau band, the occupation of electrons for $E_{\nu,s}$  $(\nu\geq1)$ bands vanishes and the $\nu=0$ Landau band is partially filled as shown in Fig.~\ref{Fig:Band}~(c). Therefore, we only need to take into account the impurity scatterings between $\nu=0$ Landau states. Using Eq. (\ref{Ur})
\begin{eqnarray}\label{U00}
U_{0,0} &\equiv& \langle 0, k_x,k_z | U(\mathbf{r}) | 0,k'_x,k'_z \rangle \nonumber\\
&=& \sum_i \mathcal{I}^{0;0}_{k_x,k_z;k_x',k_z'}(\boldsymbol{R}_i),
\end{eqnarray}
where the integral
\begin{eqnarray}\label{I00_1}
\mathcal{I}^{0;0}_{k_x,k_z;k_x',k_z'}(\boldsymbol{R}_i) &=& \int \frac{d^{3}\mathbf{r}}{L_xL_z} \psi^{0*}_{k_x,k_z}(y)\psi^0_{k'_x,k_z'}(y)
\nonumber\\
&&\times U(\mathbf{r}-\boldsymbol{R}_i)
 e^{i(k'_x-k_x)x+i(k'_z-k_z)z} \nonumber
\end{eqnarray}
describes the probability amplitude of charge carriers scattered from state $|0,k_x,k_z \rangle$ to $|0,k'_x,k'_z \rangle$ by an impurity located at position $\boldsymbol{R}_i$ in real space.  We may then write the impurity scattering potential as
\begin{eqnarray}
U(\mathbf{r}-\boldsymbol{R}_i)=\frac{1}{(2\pi)^3}\int d^{3}\boldsymbol{q} U(\boldsymbol{q}) e^{i\boldsymbol{q}\cdot(\mathbf{r}-\boldsymbol{R}_i)}
\end{eqnarray}
where the Fourier transform is given by
\begin{eqnarray}
U(\boldsymbol{q}) =\frac{e^2}{\epsilon(q^2+\kappa^2)}.
\end{eqnarray}
Substituting this into Eq.~(\ref{I00_1}) and integrating along the $x$ and $z$ directions gives
\begin{eqnarray}\label{I00_2}
\mathcal{I}_{0,0}(\boldsymbol{R}_i) &=& \int \frac{d^{3}\boldsymbol{q}}{2\pi L_xL_z} e^{-i\boldsymbol{q}\cdot \boldsymbol{R}_i} U(\boldsymbol{q}) \delta_{q_x,k'_x-k_x} \delta_{q_z,k'_z-k_z} \nonumber\\&\times&\int dy \psi^{0*}_{k_x}(y)\psi^0_{k'_x}(y)e^{iq_yy},
\end{eqnarray}
where $\delta$ is the Kronecker symbol. In terms of  $\mathcal{I}_{0,0}(\boldsymbol{R}_i)$, the absolute value squared of the scattering matrix element between the states of the $\nu=0$ Landau band can be written as
\begin{eqnarray}\label{U002}
\left| U_{0,0} \right|^2 = \sum_i \sum_j \mathcal{I}_{0,0}(\boldsymbol{R}_i)\mathcal{I}^*_{0,0}(\boldsymbol{R}_j).
\end{eqnarray}
After averaging over impurity configurations we obtain
\begin{eqnarray}
\langle\left| U_{0,0} \right|^2\rangle_{\rm imp}&=& n_{\rm imp} \int \frac{d^{3} \boldsymbol{q}}{2\pi L_xL_z} U^2(\boldsymbol{q}) \\
&\times& e^{-\frac{\ell_B^2}{2}(q_x^2+q_y^2)}\delta_{q_x,k'_x-k_x}\delta_{q_z,k'_z-k_z}\nonumber,
\end{eqnarray}
where we have used the random impurity approximation
\begin{eqnarray}
\langle \sum_{i,j}e^{i\boldsymbol{q}\cdot \boldsymbol{R}_i}e^{i\boldsymbol{q}'\cdot \boldsymbol{R}_j}\rangle_{\rm imp} \approx n_{\rm imp} (2\pi)^3 \delta_{\boldsymbol{q},-\boldsymbol{q}'},
\end{eqnarray}
where $n_{\rm imp}$ is the density of impurities over the full sample. Using the energy dispersion of the Landau bands in Eq.~(\ref{Eq:Ek}), the velocity $\hbar v_z\equiv\partial E_{\nu}/\partial k_z$ is found as
\begin{eqnarray}\label{kF-B}
\hbar v_z =
\begin{cases}
2uk_z\pm v/ \left|2vk_z-\omega\right|, & \mbox{  for $\nu \geq 1$},\\
2u \left(k_z-v/2u\right), & \mbox{  for $\nu=0$}.
\end{cases} \nonumber \\
\end{eqnarray}
Measured from the band bottom at $k_z=v/2u$, the carrier density $n$ of the $\nu=0$ Landau band is given by
\begin{eqnarray}
n &=& \frac{1}{2\pi \ell_B^2} \frac{2|k_{\mathrm{F}}^0-v/2u|}{2\pi} .
\end{eqnarray}
Combining the above two relations, we have
\begin{eqnarray}\label{vF-B}
\hbar v_{\mathrm{F}}^0=4u\pi^2\ell_B^2 n,
\end{eqnarray}
so $\hbar v_{\mathrm{F}}^0$ is proportional to $1/B$ for a fixed $n$.

The transport time ${\tau^{0,\mathrm{tr}}_{k_x,k_z}}$ of electrons in the $\nu=0$ Landau band is defined as
\begin{eqnarray}\label{TransTime}
\frac{\hbar}{\tau^{0,\mathrm{tr}}_{k_x,k_z}}&\equiv& 2 \pi \sum_{k'_x,k'_z} \langle|U_{0,0}|^2\rangle_{\rm imp}\delta\left(E_{\mathrm{F}}-E_0^{k'_z}\right)\left(1-\frac{v^z_{k'_z}}{v^z_{k_z}}\right),\nonumber\\
\end{eqnarray}
As shown in Fig.~\ref{Fig:Band}~(b), the parabolic $\nu=0$ Landau band crosses the Fermi energy at two points of wave vector, denoted as $k_z=\pm k_{\mathrm{F}}^0$, from which we can simplify Eq.~(\ref{TransTime}) as
\begin{eqnarray}\label{tau_fermi}
\frac{\hbar}{\tau^{0,\mathrm{tr}}_{k_x,k_z=\pm k_{\mathrm{F}}^0}} &=& 4\pi \Lambda \sum_{k'_x,k'_z} \langle|U_{0,0}|^2\rangle_{\rm imp}  \frac{\delta_{k_{\mathrm{F}}^0,k'_z}}{\hbar v^0_{\mathrm{F}}},
\end{eqnarray}
where $\Lambda$ is an extra correction factor introduced to avoid the van Hove singularity at the band edge of the 1D Landau band \cite{Lu15prb-QL}. Equation~(\ref{tau_fermi}) suggests that only the backscattering survives, i.e., $q_z=2k_{\mathrm{F}}^0$. Substituting Eq.~(\ref{U002}) into Eq.~(\ref{tau_fermi}), and considering $\ell_B^2 \kappa^2 \ll 1$ in strong magnetic fields, we have
\begin{eqnarray}
\frac{\hbar}{\tau^{0,\mathrm{tr}}_{k_x,k_z=\pm k_{\mathrm{F}}^0}} &=& n_{\rm imp} \frac{\Lambda}{\hbar v_{\mathrm{F}}^0} \frac{e^4}{\epsilon^2} \frac{1}{2\pi \kappa^2}.
\end{eqnarray}
By using Eq.~(\ref{kappa}), we arrive at the transport time of the $\nu=0$ band in strong magnetic fields
\begin{eqnarray}\label{tau_res}
\frac{\hbar}{\tau^{0,\mathrm{tr}}_{k_x,k_z=\pm k_{\mathrm{F}}^0}}
&=& \frac{\pi n_{\rm imp} e^2 \ell_B^2 \Lambda}{\epsilon}.
\end{eqnarray}

\section{Longitudinal Magnetoconductivity} \label{Sec:Conductivity}
With the transport time, we are ready to calculate the longitudinal conductivity of the $\nu=0$ band. Along the $z$-direction, the semiclassical Drude conductivity can be calculated as
\begin{eqnarray}
\sigma_{zz,0}=\frac{e^2\hbar}{2\pi V}\sum_{k_x,k_z}(v^z_0)^2G^{\mathrm{R}}_0G^{\mathrm{A}}_0,
\end{eqnarray}
where $e$ is the electron charge, $V=L_xL_yL_z$ is the sample volume, $v^z_0={\partial E^0_{k_z}}/{\hbar\partial k_z}$ is the velocity along the $z$-direction for a state with wave vector $k_z$ in the $\nu = 0$ Landau band,
\begin{eqnarray}
G^{\mathrm{R}/\mathrm{A}}=\frac{1}{E_{\mathrm{F}}-E^0_{k_z}\pm i\hbar/{2\tau_{k_x,k_z}^0}}
\end{eqnarray}
are the retarded/advanced Green's functions with $\tau_{k_z,k_x}^0$ the lifetime of a state with wave vector $k_x$ and $k_z$ in the $\nu=0$ Landau band. In the diffusive regime, $G^{\mathrm{R}}_0G^{\mathrm{A}}_0$ can be replaced by
\begin{eqnarray}\label{G2}
G^{\mathrm{R}}_0G^{\mathrm{A}}_0=\frac{2\pi}{\hbar}\tau^{0}_{k_x,k_z}\delta\left(E_{\mathrm{F}}-E^0_{k_z}\right)
\end{eqnarray}
Changing the summations into integrals through
\begin{eqnarray}
\sum_{k_z}\longrightarrow L_z\int\frac{dk_z}{2\pi},\ \  \sum_{k_x}\longrightarrow {L_x} \int^{L_y/2\ell_B^2}_{-L_y/2\ell_B^2}\frac{dk_x}{2\pi},
\end{eqnarray}
where the $k_z$ integral covers the entire Brillouin zone and the $k_x$ integral is confined by the degeneracy of the Landau levels, the conductivity can be expressed as
\begin{eqnarray}\label{con}
\sigma_{zz,0}=\frac{e^2\hbar}{2\pi L_y}
\int \frac{dk_x}{2\pi} \int\frac{dk_z}{2\pi}(v^z_0)^2G^{\mathrm{R}}_0G^{\mathrm{A}}_0.
\end{eqnarray}
By using Eq.~(\ref{G2}), we have
\begin{eqnarray}\label{c1}
\sigma_{zz,0}=\frac{e^2v^0_{\mathrm{F}} \Lambda}{\pi\hbar L_y}\int^{L_y/2\ell_B^2}_{-L_y/2\ell_B^2} \frac{dk_x}{2\pi} \tau^{0,\mathrm{tr}}_{k_x,k_z=k^0_{\mathrm{F}}},
\end{eqnarray}
where the Landau degeneracy has already been taken into account. Now using
the transport time in Eq.~(\ref{tau_res}), we finally arrive at
\begin{eqnarray}\label{Result}
\sigma_{zz,0}= \frac{e^2}{h}\frac{\epsilon \hbar v_{\mathrm{F}}^0}{\pi^2 n_{\rm imp}e^2 \ell_B^4}.
\end{eqnarray}

 Considering  the $1/\ell_B^4\propto B^2$ in Eq.~(\ref{Result}) and the velocity along z-direction obtained in Eq.~(\ref{vF-B}), $\sigma_{zz,0}$ shows a positive linear dependence on magnetic field, i.e.,
\begin{eqnarray}
\sigma_{zz,0} \propto B.
\end{eqnarray}
In a parallel magnetic field, there is no Hall effect, thus the resistivity $\rho_{zz,0}$ is the inverse of the conductivity. Hence,
\begin{eqnarray}
\rho_{zz,0} \propto \frac{1}{B},
\end{eqnarray}
which means that the resistivity drops with increasing magnetic field. In other words, we find a negative longitudinal magnetoresistance in the quantum limit of the Kramers Weyl cone in the presence of a screened Coulomb impurity scattering potential.

\section{Gaussian scattering potential}\label{Sec:Gaussian}

Another common choice of the scattering potential is the Gaussian one. We can show that there is no negative magnetoresistance in the presence of the Gaussian scattering potential in the quantum limit of the Kramers Weyl cone. The Hamiltonian of the Gaussian scattering reads \cite{Goswami15prb,ZhangSB16njp}
\begin{eqnarray}\label{U-Gaussian}
U(\mathbf{r}) &=& \sum_i \frac{u_i}{(d\sqrt{2\pi})^3}  e^{-|\mathbf{r}-\boldsymbol{R}_i|^2/2d^2},
\end{eqnarray}
or in terms of its Fourier transform
\begin{eqnarray}
U(\mathbf{r})=\sum_i \frac{1}{(2\pi)^3}\int d^{3}\boldsymbol{q} U(\boldsymbol{q}) e^{i\boldsymbol{q}\cdot(\mathbf{r}-\boldsymbol{R}_i)}
\end{eqnarray}
where
\begin{eqnarray}
U(\boldsymbol{q}) &=& u_i  e^{-q^2d^2/2},
\end{eqnarray}
and $u_i$ measures the scattering strength at $\boldsymbol{R}_i$. A major difference here is that the range of the potential $d$ is not a function of the magnetic field. As $d$ shrinks to zero, the potential reduces to a delta potential.

Following the same procedure that was used to obtain Eq.~(\ref{tau_res}), the transport time
for the Gaussian scattering potential is found to be
\begin{eqnarray}\label{tau-tr-gauss-twonode-B}
\frac{\hbar}{\tau^{0,0,\rm G}_{k_x,k_z=k_{\mathrm{F}}^0}}
= \frac{\Lambda V_{\text{imp}}}{\hbar v_{\mathrm{F}}^0}
\frac{ e^{-2(2d^2+\ell_B^2) (k_{\mathrm{F}}^0)^2}}{ \pi ( 2d^2+\ell_B^2 )}.
\end{eqnarray}
Inserting Eq.~(\ref{tau-tr-gauss-twonode-B}) into~(\ref{c1}), one obtains the conductivity
\begin{eqnarray}
\sigma_{zz,0}^{\rm G}=\frac{e^2}{h}
\frac{(\hbar v_{\mathrm{F}}^0)^2( 2d^2+\ell_B^2 )}{V_{\text{imp}}\ell_B^2 }e^{2(2d^2+\ell_B^2) (k_{\mathrm{F}}^0)^2}.
\end{eqnarray}
According to Eqs.~(\ref{kF-B}) and~(\ref{vF-B}), $k_{\mathrm{F}}^0\propto v_{\mathrm{F}}^0\propto 1/B$, so $\sigma_{zz,0}^{\rm G}$ will decrease with increasing $B$, which cannot give a negative magnetoresistance. We conclude that the screened Coulomb scattering potential is an essential ingredient for a negative magnetoresistance.

\section{Anisotropic Kramers Weyl cone}\label{Sec:Anisotropy}

\begin{figure}
\centering
\includegraphics[width=0.45\textwidth]{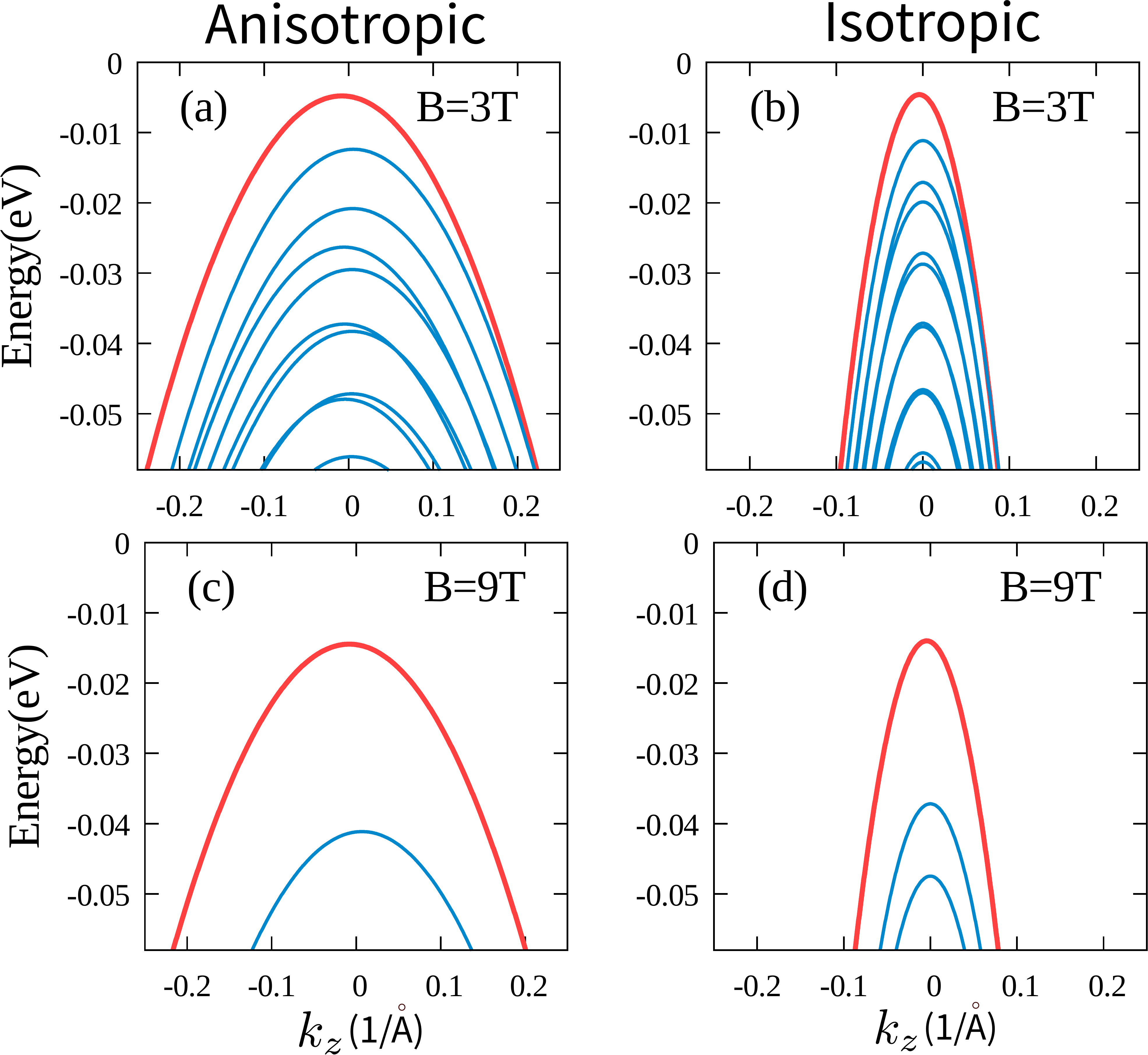}
\caption{The comparison of the Landau bands for an anisotropic Kramers Weyl cone and an isotropic Weyl cone at different magnetic fields. The red curves are the $\nu=0$ Landau bands. The blue curves are Landau bands with $\nu\ge 1$. The parameters for the anisotropic case are $u_x=-14.8$ eV\AA$^2$, $u_y=-2.97$ eV\AA$^2$, $u_z=-1.55$ eV\AA$^2$, $v_x=0.079$ eV\AA, $v_y=0.066$ eV\AA, $v_z=0.020$ eV\AA. The parameters for the isotropic case are $u_x=u_y=u_z=-6.44$ eV\AA$^2$, $v_x=v_y=v_z=0.055$ eV\AA.}
\label{Fig:Anisotropic-Landau}
\end{figure}

The original model for the Kramers Weyl cone in $\beta$-Ag$_2$Se is anisotropic. Now we show that the anisotropy does not qualitatively change the negative magnetoresistance.
The original model reads \cite{ZhangCL17prb_Ag2Se}
\begin{eqnarray} \label{Eq:anisotropic-model}
\mathcal{H}=\sum_{i=x,y,z} (u_i k_i^2 + v_i k_i\sigma_i),
\end{eqnarray}
where the model parameters $u_x=-14.8$ eV\AA$^2$, $u_y=-2.97$ eV\AA$^2$, $u_z=-1.55$ eV\AA$^2$, $v_x=0.079$ eV\AA, $v_y=0.066$ eV\AA, $v_z=0.020$ eV\AA. Because of the anisotropy, there is no analytical solution for the Landau bands. We numerically solve the energy spectrum of the Landau bands. Figure~\ref{Fig:Anisotropic-Landau} shows that the structure of the Landau spectrum does not change qualitatively in the presence of the anisotropy, so one can also expect that a negative magnetoresistance is obtained in the anisotropic case. This justifies our calculation using the isotropic model.

\section{Conclusion and Discussion}\label{Sec:Discussion}

In conclusion, we presented a theory for the negative magnetoresistance observed in the quantum limit of  $\beta$-Ag$_2$Se, a paradigmatic Kramers-Weyl semimetal, in parallel magnetic fields \cite{ZhangCL17prb_Ag2Se}.
It requires several ingredients: (1) a Kramers Weyl cone; (2) impurities with a screened Coulomb potential; (3) a fixed carrier density; (4) the quantum limit under parallel magnetic fields. These conditions naturally exist in $\beta$-Ag$_2$Se.

{ The negative magnetoresistance has been previously studied in the quantum limit of the band-inversion Weyl semimetals in parallel magnetic fields \cite{Lu15prb-QL,Goswami15prb,ZhangSB16njp}. In the presence of the charge-neutral Gaussian potential, the conductivity is linearly proportional to the magnetic field \cite{Goswami15prb,ZhangSB16njp}. In the ionic potential under the Thomas-Fermi approximation, the conductivity is proportional to $B^2$ in all cases \cite{Goswami15prb}. The novel mechanism for the negative magnetoresistance that we present in this work has not been covered in these previous studies.
In particular, here the negative magnetoresistance is not related to the chiral anomaly, which has been used to explain the effect in band-inversion induced Weyl semimetals.}

In band-inversion Weyl semimetals, two Weyl cones of opposite chirality have to appear in pairs. In a strong magnetic field, the $\nu=0$ Landau bands from the two cones have opposite velocities inherited from the chirality of the Weyl cones.
In contrast, in Kramers Weyl semimetals, the energy difference between Kramers Weyl nodes of opposite chirality can be very large, because, e.g., one is located at the $\Gamma$ point and the other at the Brillouin zone corner.
As the Fermi energy crosses the vicinity of one of the Weyl nodes, the contributions to transport coming from the other Weyl node can be safely ignored. For instance, the $\nu=0$ band of only one Weyl node in Fig.~\ref{Fig:Band}~(b) has no sense of the inherited chirality, although it looks similar to the quantum limit of the band-inversion Weyl semimetal. Therefore, a negative magnetoresistance of the single Kramers Weyl cone has nothing to do with the chiral anomaly.

According to the above result for the hole carriers, the key ingredients of the negative magnetoresistance are the parabolic dispersion of the 0th Landau band and the screened Coulomb scattering potential. According to the DFT band structure in Ref. [26], there is another electron pocket on the Fermi surface and its dispersion looks quite conventional. The conventional electron pocket is expected to have the parabolic dispersion for the 0th Landau band. Therefore, along with the same screened Coulomb scattering potential, the electron pocket is also expected to give the negative magnetoresistance.

\begin{acknowledgments} We thank helpful discussions with Cheng-Long Zhang, Su-Yang Xu, and Shuang Jia. This work was supported by Guangdong Innovative and Entrepreneurial Research Team Program (Grant No. 2016ZT06D348), the National Key R \& D Program (Grant No. 2016YFA0301700), the
National Natural Science Foundation of China (Grants No. 11525417 and No. 11574127), and the Science, Technology and Innovation Commission of Shenzhen Municipality (Grant No. ZDSYS20170303165926217 and JCYJ20170412152620376).
F.S. and T.N. acknowledge support from the Swiss National Science Foundation (Grant No. 200021\_169061) and from the European Unions Horizon 2020 research and innovation program (ERC-StG-Neupert-757867-PARATOP). X.W. also acknowledges support by the Fundamental Research Funds for the Central Universities (No. 020414380085).
\end{acknowledgments}


%

\end{document}